# Purification of total flavonoids from *Aurea Helianthus* flowers and In Vitro Hypolipidemic Effect

Hyon-il Ri[1], Chol-song Kim[2,3], Un-hak Pak[1], Myong-su Kang[2], Tae-mun Kim[2*]

1. Kim Hyong Jik University of Education，Pyongyang 999093，Democratic People's Republic of Korea
2. Kim Il Sung University, Pyongyang 999095，Democratic People's Republic of Korea
3. Peking University, Beijing 100871, PR China

**A B S T R A C T**

The effects of purification methods and its hypolipidemic function on the total flavonoids of *Aurea Helianthus* flower were investigated. Liquid-liquid extraction of ethanol extract from *Aurea Helianthus* flower was carried out by using different polar solvents. The extract with the highest total flavonoid content was selected, and the optimal conditions for purification of total flavonoids were determined by purification with macroporous resin. The human digestive environment was simulated in vitro, and the binding ability of different flavonoid samples to three kinds of cholate was compared. The results showed that the purity of total flavonoids in ethanol extract was 27.8%, the purity of total flavonoids in ethyl acetate extract was 46.4%, and the purity was increased by 18.6%. Subsequent purification with AB-8 macroporous resin; loading of total flavonoids at a concentration of 5.5 mg/mL, flow rate of 1.5 mL/min, 110 mL; use of 75% ethanol, 80 mL as eluent at a flow rate of 1.5 mL The elution at /min resulted in a total flavonoid purity of 83.5 % and an increase of 37.1%, and a good purification effect was obtained. The binding rate of total flavonoids purified by AB-8 macroporous resin to sodium taurocholate, sodium glycocholate and sodium cholate was 88.2%, 73.2% and 75.8 %, respectively. The binding ability was the strongest, and the others were ethyl acetate. Extract, ethanol extract. The purity of total flavonoids showed a good correlation with the binding capacity of cholate, and the correlation coefficient was between 0.963 and 0.988. The total flavonoids of *Aurea Helianthus* flower have good bile acid binding ability and can be used as the focus of natural hypolipidemic substances.

## 1. Introduction

Aurea Helianthus, also known as Jin Furong, Vegetable Hibiscus, belonging to Malvaceae, Okra, annual herb, has the most edible, medicinal, health care and higher commercial value in more than 200 kinds of okra plants. The study found that the geranium is rich in flavonoids, which has good antioxidation, hypolipidemic, hypoglycemic, antitumor, analgesic, antipyretic, antiinflammatory, immune regulation, antiaging, liver protection and other functions, so the development of Aurea Helianthus flavonoids are particularly important [1]. In recent years, the extraction, separation and purification methods of flavonoids mainly include organic solvent reflux method, soxhlet extraction method, auxiliary extraction method, enzymatic hydrolysis method and resin method [2-4], especially solvent extraction method and macroporous resin. The method has the advantages of good purification effect and is widely used for the purification of flavonoids [5,6]. Due to the short discovery time of flavonoids in the geranium, the extraction and separation methods are not perfect enough, and the research on its physiological activity is not enough. Hyperlipidemia leads to important risk factors for cardiovascular and cerebrovascular diseases such as atherosclerosis, coronary heart disease and stroke. The incidence rate is increasing year by year, and the population is increasing gradually, which is harmful to human health. At present, the main treatment of hyperlipidemia is mainly chemical drugs, but long-term use may damage side effects such as liver and kidney [7]. Therefore, the development of natural functional components that effectively regulate blood lipids has become a hot spot. Studies have shown that flavonoids have a good lipid-lowering effect and are an important natural chemical constituent with broad development and application prospects [8-11]. However, the experimental study on the separation and purification of total flavonoids from Aurea Helianthus and its antihyperlipidemic effect in vitro has not been reported [12,13]. In this study, the solvent extraction method and macroporous resin method were used to separate and purify the total flavonoids of Aurea Helianthus flower, and the optimal conditions for the purification of total flavonoids were determined. Secondly, the total flavonoids of the flowering sunflower were used to determine the bile salt binding rate. The evaluation of blood lipid capacity provides a theoretical basis for the development of natural hypolipidemic drugs, and also lays a foundation for the development and application of Aurea Helianthus.

## 2. Materials and methods

*2.1 Materials and Instruments*

Aurea Helianthus Flower : Pyongyang(Korea); Rutin: Shanghai Shifeng Biotechnology Co., Ltd.; sodium taurocholate, sodium glycocholate, sodium cholate: Shanghai Maclean Biotechnology Co., Ltd.; pepsin (1:30) 000), trypsin (1:250): Shanghai Yuanye Biotechnology Co., Ltd.; AB-8 macroporous resin: Zhangzhou Baoen Adsorption Material Technology Co., Ltd.; sodium nitrite, aluminum nitrate, sodium hydroxide, hydrochloric acid, anhydrous Ethanol is of analytical grade: produced by Tianjin Jinhai Chemical Co., Ltd.; other reagents are of analytical grade.
UV spectrophotometer UV2450: Shimadzu Corporation, Japan; KQ-100B ultrasonic cleaner: Kunshan Ultrasonic Instrument Co.,Ltd.; RE-52AA rotary evaporator: Shanghai Yarong Biochemical Instrument Factory; Electronic analytical balance BT 255: German Sartorius ;DZF- 6020 vacuum drying oven: Shanghai Yiheng Experimental Instrument Co., Ltd.; TDL-40B desktop centrifuge: Shanghai Anting Scientific Instrument Factory.

*2.2 methods*

2.2.1 Extraction of total flavonoids from Aurea Helianthus flower

The method of VIACAVA et al. [14] was slightly modified. 20.0 g of dried Aurea Helianthus flower powder was weighed, passed through a 80 mesh sieve, and degreased with petroleum ether at 60 ° C. After evaporation, according to the ratio of 1:30 (g/mL), 75% ethanol was added, and the ultrasonic time was 30 min. A total of three times, the extracts were combined and concentrated under vacuum at 40 ° C to obtain total flavonoid extract, dried at 45 ° C for 20 h, and the crude extract of flavonoids was used for use.

2.2.2 Aurea Helianthus flower crude flavonoids with different polar solvent extraction

The method of Yun Cheng yue et al [4] was slightly modified. Weigh a certain amount of Aurea Helianthus flower extract, add water to dissolve, shake, add a certain amount of petroleum ether to shake and let stand, after the solution is completely layered, pour out the upper petroleum ether phase; the lower aqueous solution according to the above method The extraction is further carried out 4 times; the remaining aqueous phase is sequentially extracted with ethyl acetate and n-butanol according to the above steps, and the petroleum ether extract, the ethyl acetate extract, the n-butanol extract and the final aqueous phase solution are successively obtained, which will be the same. The extract is combined and concentrated under reduced pressure to obtain an extract extract which is then placed in a desiccator. The total flavonoids content in each extract was determined, and the extract with the highest total flavonoid content was selected and subsequently used for macroporous resin purification.

2.2.3 Determination of total flavonoids

The method of VIACAVA et al[14] was slightly modified. Weigh 4.20 mg of rutin standard, place it in a 25 mL volumetric flask, dissolve it in 60% ethanol and dilute to the mark and shake it to obtain the initial solution of rutin standard. Accurately draw 0.5, 1.0, 1.5, 2.0, 2.5, 3.0 mL standard solution into a 10 mL volumetric flask, dilute to 5.0 mL with 60% ethanol, add 0.4 mL of 5 % $NaNO_2$ and shake for 6 min. Add 0.4 mL of 10% $Al(NO_3)_3$ and shake well. Add 0.4 mL of 4% NaOH, finally add 10 mL of deionized water, and measure the absorbance at λ = 510 nm. Obtain the regression equation Y=4.0361X-0.1062 ($R^2$) =0.9989), the absorbance of the measured sample is substituted into the equation to calculate the total flavonoid content in the sample.

2.2.4 Macroporous resin pretreatment

Take a certain amount of macroporous resin soaked in 95% ethanol for 10 h, fully swell and rinse with deionized water until tasteless, then soaked in 1% hydrochloric acid for 10 h, washed with distilled water until neutral, and then soaked with 5% sodium hydroxide. h, wash with distilled water until neutral, and dry for use.

2.2.5 Static test

Weighed 3.0 g of the above-mentioned 8 kinds of macroporous resins, and added 50.0 mL of a sample of 3.50 mg/mL of flavonoids in a 250 mL stoppered flask to shake (30 °C). , 150 r/min, 20 h), the macroporous resin was filtered, the absorbance of the filtrate was measured, and the equilibrium mass concentration was calculated. Rinse the above-mentioned saturated resin with distilled water for 2 to 3 times, add 80.0 mL of ethanol with a volume fraction of 95% (30 °C, 150 r/min, 20 h), measure the absorbance of each resin desorption solution and calculate the mass concentration of the desorbed solution by filtration. The adsorption amount, adsorption rate, desorption amount and desorption rate of each resin were calculated according to the following formula, and the best macroporous resin was selected.

$$\text{Adsorption amount} = \frac{C_1 - C_2}{m} \times V_1; \quad \text{adsorption rate} = \frac{C_1 - C_2}{C_1} \times 100\%;$$

$$\text{desorption amount} = \frac{C_3 V_2}{m}; \quad \text{desorption rate} = \frac{C_3 V_2}{(C_1 - C_2) V_1} \times 100\%。$$

Where: m is the mass of the resin (g); $C_1$ is the mass concentration of total flavonoids in the adsorbent (mg/mL); $C_2$ is the mass concentration of total flavonoids in the equilibrium solution (mg/mL); $C_3$ is the mass concentration of total flavonoids in the desorption solution (mg/ mL); $V_1$ is the volume of the adsorbent (mL); $V_2$ is the volume of the desorbed solution (mL).

2.2.6 Dynamic test

The method of Jin Huiming et al. [15] was slightly modified. After pretreatment of the macroporous resin selected by the static test, a certain amount of macroporous resin was weighed and packed in a wet manner (Φ 18 mm × 300 mm). The total flavonoid solution of the golden flower sunflower is applied to the column. After the total flavonoid solution passes through the resin column, the resin column is washed with about 80-100 mL of distilled water, and finally eluted with an ethanol solution. By measuring the mass concentration of total flavonoids in the adsorption solution and the eluent and calculating the adsorption rate, the mass concentration of total flavonoids in the sample solution, the loading flow rate, the volume of the sample solution, the number of elution liquids, the elution flow rate, and the volume of the eluent were investigated. The optimum conditions for the purification of total flavonoids were determined by the effect on the properties of macroporous resins.

2.2.7 Determination of the purity of total flavonoids

The ethanol extract, the solvent extract of different polar solvents and the macroporous resin purification solution were respectively concentrated in a rotary evaporator to form a dipstick, and dried in a vacuum freeze dryer for 24 h. Accurately weigh a certain amount with 95% ethanol solution, determine the total flavonoid concentration in the solution by the method of Section 1.2.3, and calculate the purity according to the following formula:

Total flavonoid purity = CV / m × 100%. Where: C is the total flavonoid concentration (mg/mL); V is the sample volume (mL); m is the flavonoid dry powder mass (mg).

2.2.8 Determination of in vitro combined cholate ability of total flavonoids

2.2.8.1 Drawing of the standard curve of cholate

The method used in YU Meihui et al. [16] has been slightly modified. Weigh a certain amount of sodium cholate, sodium glycocholate and sodium taurocholate, dissolved in phosphate buffer solution (pH 6.3) and formulated into different concentrations (0.10, 0.15, 0.20, 0.25, 0.30, 0.35 mmol / L) Standard solution; accurately absorb 2.5 mL of each standard solution, then add 7.5 mL of 60% sulfuric acid, shake well, and then bathe at 70 °C for 20 min, remove the ice bath for 5 min, measure the absorbance at 387 nm, and take the cholate content as the abscissa. The absorbance is plotted on the ordinate to obtain a standard curve of cholate content, and the concentration of cholate in the sample solution is determined from the standard curve.

2.2.8.2 Determination of the binding capacity of cholate

The method of Li Dehai et al [17] was slightly modified. Weigh 10 mg of different flavonoids in a 50 mL stoppered flask, add 1 mL of 10 mg/mL pepsin (prepared with 0.1 mol/L phosphate buffer of p H 6.3), and 1 mL of 0.01 mol/L HCl solution. Simulate the stomach environment and digest it at 37 °C for 1 h. Adjust the p H value to 6.3 with 0.1 mol/L sodium hydroxide solution, then add 4 mL 10 mg/mL trypsin (0.1 mol/p H 6.3) Prepared by L-phosphate buffer), digested at 37 °C for 1 h at a constant temperature, and simulated the intestinal environment for digestion. Each sample was supplemented with 4 mL of 1 mmol/L sodium cholate, sodium glycocholate and sodium taurocholate (prepared with 0.1 mol/L phosphate buffer of p H 6.3), and the mixture was shaken at 37 °C for 1 h. Move to a centrifuge tube, centrifuge at 4000 r/min for 20 min, determine the concentration of cholate in the supernatant, and calculate the cholate binding rate according to the following formula:

$$\text{Cholate salt binding rate} = \frac{C_4 - C_5}{C_4} \times 100\%。$$

Where ; $C_4$ is the amount of cholate added (μmol); $C_5$ is the remaining amount of cholate (μmol).

2.2.9 Data Analysis

All experimental data are the average of the results of 3 replicate experiments. Statistical analysis was performed using Origin 8.0 and SPSS Statistics 22 software.

# 3. Results and discussion

*2.1 Extraction results of different polar solvents of flavonoids*

According to the methods of 1.2.1 and 1.2.2, the quality of the different polar components of the total flavonoids of the sunflower flower and the purity of the total flavonoids were as shown in Table 1.

**Table 1**. Extraction effect of different polar solvents of flavonoids from *Aurea Helianthus* flower

|  | ethanol | petroleum ether | ethyl acetate | n-butanol | water |
| --- | --- | --- | --- | --- | --- |
| quality (g) | 9.63±0.05[b] | 0.48±0.03[d] | 2.31±0.05[a] | 1.18±0.03[b] | 5.62±0.04[a] |
| total flavonoid purity (%) | 27.8±0.4[c] | 8.1±0.2[a] | 46.4±0.5[c] | 38.5±0.4[d] | 18.5±0.3[c] |

The values in the table are expressed as mean ± sd, 3 replicates; different superscripts indicate significance level ($p < 0.05$).

It can be seen from Table 1 that 9.63 g of ethanol extract was obtained from 20.0 g of dried Aurea Helianthus flower, and it was separated into petroleum ether phase 0.48 g, ethyl acetate phase 2.31 g, n-butanol phase 1.18 g, and water phase 5.62 by solvent extraction with different polar solvents. g. The purity of total flavonoids in each extract phase was different. The purity of total flavonoids in ethyl acetate phase was significantly higher than that in other phases (p<0.05), and its purity increased from 27.8 % to 46.4%, an increase of 18.6%. Therefore, the total flavonoids purification ratio in the ethyl acetate extract phase was 1.67 times, which was subsequently used for macroporous resin purification.

*2.2 static test results*

The static adsorption and desorption capacities of the eight macroporous resins of the total flavonoids of Aurea Helianthus flower are shown in Table 2.

It can be seen from Table 2 that the adsorption capacity and adsorption rate of the eight macroporous resins are in the order of AB-8 > HPD700 > HPD500 > D101 > HPD600 > HPD200 > S-8 > X-5, of which AB-8 macroporous resin The adsorption capacity and adsorption rate were 35.15 mg/g and 60.25%, respectively. The order of 8 macroporous resin desorption rates is HPD200 > HPD500 > AB-8 > HPD700 > S-8 > HPD600 > X-5> D101.

Table 2. Result of static adsorption and desorption of eight types of resin

| Resin model | Adsorption amount (mg/g) | Adsorption rate (%) | Desorption amount (mg/g) | Desorption rate (%) |
|---|---|---|---|---|
| HPD–700 | 32.36±0.23 | 55.47±0.28 | 22.12±0.21 | 68.36±0.33 |
| HPD–600 | 28.32±0.25 | 48.54±0.26 | 18.55±0.16 | 65.52±0.31 |
| HPD–500 | 31.42±0.23 | 53.86±0.32 | 23.94±0.23 | 76.21±0.27 |
| HPD–200 | 26.85±0.16 | 46.02±0.41 | 20.98±0.31 | 78.13±0.23 |
| X-5 | 22.32±0.13 | 38.26±0.25 | 14.09±0.18 | 63.16±0.12 |
| D101 | 30.24±0.28 | 51.84±0.31 | 18.66±0.25 | 61.71±0.35 |
| AB-8 | 35.15±0.28 | 60.25±0.23 | 26.06±0.36 | 74.14±0.27 |
| S -8 | 26.13±0.12 | 44.79±0.14 | 17.79±0.18 | 68.09±0.22 |

The values in the table are expressed as mean ± standard deviation, 3 repeated experiments

It can be seen from Table 2 that the adsorption capacity and adsorption rate of the eight macroporous resins are in the order of AB-8 > HPD700 > HPD500 > D101 > HPD600 > HPD200 > S-8 > X-5, of which AB-8 macroporous resin The adsorption capacity and adsorption rate were 35.15 mg/g and 60.25%, respectively. The order of 8 macroporous resin desorption rates is HPD200 > HPD500 > AB-8 > HPD700 > S-8 > HPD600 > X-5> D101. The desorption rate of HPD200 was 78.13%, the highest, followed by HPD500, AB-8, and the desorption rate was 76.21% and 74.14%, respectively. Considering the separation and enrichment effect of total flavonoids of Golden Sunflower, the ease of elution and the desorption rate, AB-8 resin was selected as the purified macroporous resin.

*2.3 Dynamic test results*

2.3.1 Determination of the total concentration of total flavonoids in the sample solution

The effect of the concentration of total flavonoids in the sample solution on the adsorption effect is shown in Figure 1.

It can be seen from Fig. 1 that the concentration of total flavonoids in the sample solution is in the range of 2.5-5.5 mg/mL, and the adsorption rate increases with the increase of the mass concentration. When the concentration of total flavonoids in the sample solution is 5.5 mg/mL, the adsorption rate is 57.5. %, which reaches the maximum, and then decreases as the mass concentration increases. Therefore, it was determined that the total flavonoid concentration of the sample solution was 5.5 mg/mL.

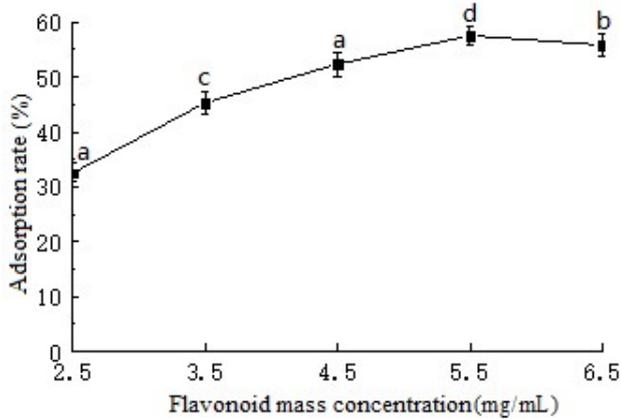
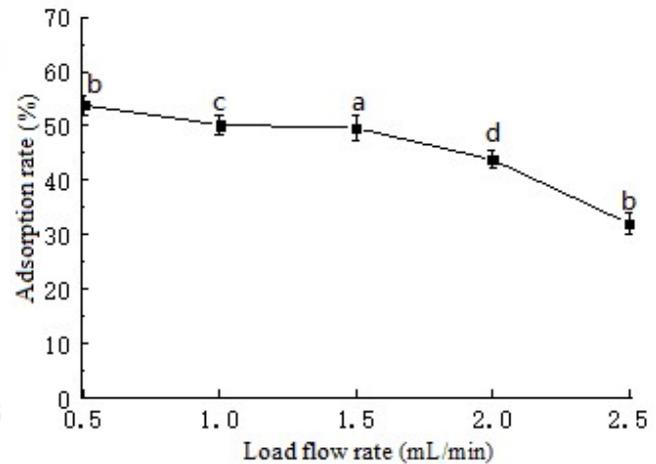

Fig. 1　Effect of total flavonoids concentration in sample on the adsorption efficiency

Fig. 2　Effect of sample flow rate on the adsorption efficiency

2.3.2 Effect of loading flow rate on adsorption effect

The effect of the loading flow rate on the adsorption effect is shown in Figure 2. It can be seen from Fig. 2 that the adsorption rate decreases with the increase of the loading flow rate in the range of 0.5-1.5 mg/mL, and the difference is not significant (p>0.05). The loading rate can continue to increase, and the adsorption rate is obvious. Decrease. Therefore, considering the two factors of adsorption effect and work efficiency, it is determined that the sample flow rate is 1.5 mL/min.

2.3.3 Effect of the volume of the sample solution on the adsorption effect

The effect of the volume of the sample solution on the adsorption effect is shown in Figure 3.

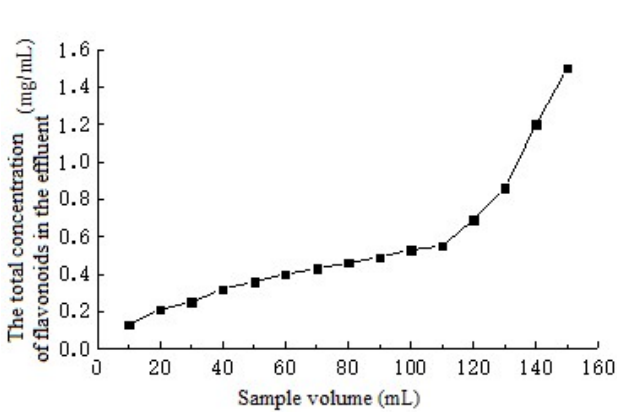
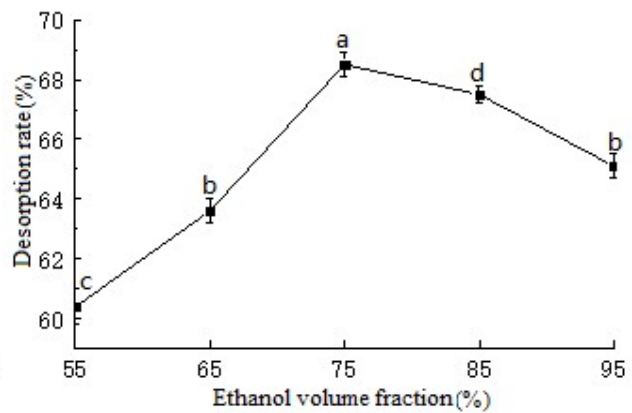

Fig. 3　Dynamic adsorption curve

Fig. 4　Effect of eluent concentration on the desorption efficiency

It can be seen from Fig. 3 that the mass concentration of total flavonoids in the effluent increases with the volume of the sample solution. When the volume of the sample solution reaches 120 mL, the total concentration of flavonoids in the effluent is 0.62 mg/mL, which has exceeded the total flavonoids in the sample. The mass concentration is 1/10 [18], so 120 mL is considered as the leak point. Therefore, in order to avoid waste of the flavonoid solution, it is determined that the volume of the sample solution is 110 mL.

2.3.4 Effect of the number of elution liquids on the desorption effect

The effect of the number of elution liquids on the desorption effect is shown in Figure 4. It can be seen from Fig. 4 that the ethanol volume fraction is 55 % to 75%, and the desorption rate increases with the increase of the ethanol volume fraction; when the ethanol volume fraction is 75%, the desorption rate is 68.5 %, the highest; then the ethanol volume fraction is increased. The desorption rate is reduced. Therefore, it is determined that 75% is the optimum eluent ethanol volume fraction.

### 2.3.5 Effect of elution flow rate on desorption

The effect of the elution flow rate on the desorption effect is shown in Figure 5. It can be seen from Fig. 5 that when the elution flow rate is 0.5-1 mL/min, the desorption rate increases gradually with the increase of the elution flow rate, and the desorption rate decreases gradually when the elution flow rate is 1.0-2.5 mL/min. When the elution flow rate was 1 mL/min and 1.5 mL/min, the desorption rates were 67.6 % and 67.2 %, respectively, and the difference was not significant ( $p > 0.05$ ). Therefore, considering the desorption effect and work efficiency, the elution flow rate was determined to be 1.5 mL/min.

### 2.3.6 Effect of eluent volume on desorption

The effect of the volume of the eluent on the desorption effect is shown in Figure 6. It can be seen from Fig. 6 that the mass concentration of total flavonoids eluted significantly increases with the volume of the eluent; when the eluent volume is 40 mL (the fourth tube), the total flavonoid concentration is 9.82 mg/mL. Reaches the maximum value. After increasing the volume of the eluate, the mass concentration of the total flavonoids eluted decreased. When the eluent volume reached 90 mL (the 9th tube), the total flavonoid concentration was only 0.02 mg/mL. Therefore, it is determined that 80 mL (the eighth tube) is the optimum eluent volume.

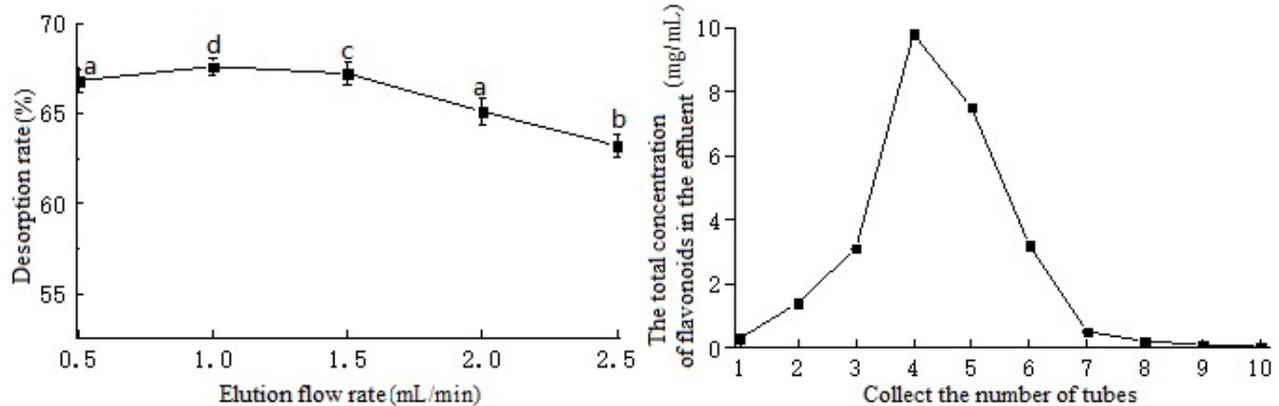
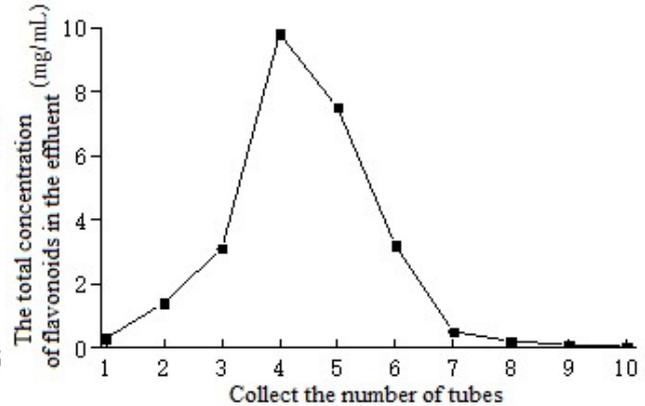

Fig. 5 Effect of elution velocity on the desorption efficiency  Fig. 6 Dynamic desorption curve

### 2.3.7 Determination of the purity of total flavonoids

The purity of total flavonoids was calculated according to the formula in Section 1.2.7 as shown in Figure 7. It can be seen from Fig. 7 that the total flavonoids of the sunflower flower are extracted by different polar solvents, and have a certain concentration and enrichment effect after purification by AB-8 macroporous resin. The purity of the product after purification by AB-8 macroporous resin is 83.5 %. From 46.4% of the ethyl acetate extract, it increased by 37.1%; by 27.8% of the ethanol extract, it increased by 55.7 %. From the overall analysis, the total flavonoids of Camellia sinensis were extracted with different polar solvents and purified with AB-8 macroporous resin, which showed good purification effect, indicating that this method has certain feasibility and can effectively improve the total sunflower flower Flavonoid purity.

*2.4 Determination of the ability of flavonoids in vitro*

### 2.4.1 Chethate standard curve

The regression curve equation of sodium cholate is *Y=0.2832X-0.2518 (R2=0.996)*, and the regression curve equation of sodium taurocholate is *Y=3.6251X+0.3302 (R2=0.992)*, the regression curve of sodium glycocholate The equation is *Y=1.9023X+0.6203 (R2=0.995)*, and the linear curves of the three standard curves are good, and the concentration of cholate can be calculated.

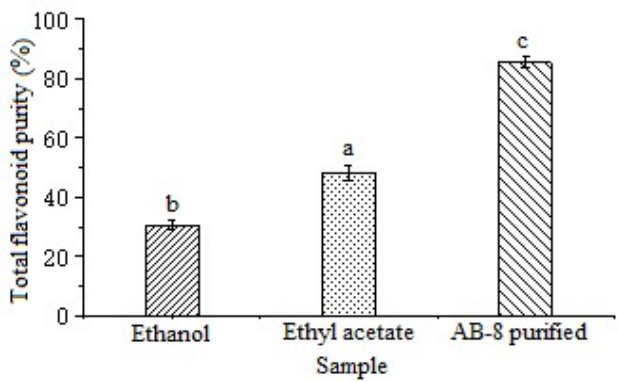 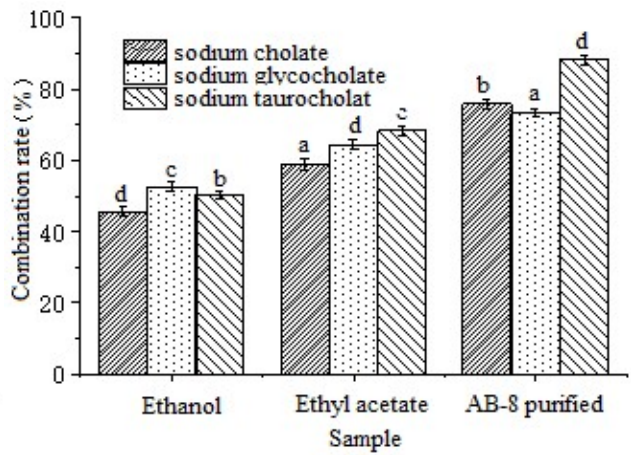

Fig.7 The effect of the purification method on the purity of total flavonoids

Fig.8 The effect of flavonoids on the binding rate of cholinate

2.4.2 Determination of the binding capacity of cholate

The effect of total flavonoid samples on bile salt binding rate is shown in Figure 8. It can be seen from Fig. 8 that the total flavonoids of the flower geranium has different obvious binding ability to the three kinds of cholates, and the order from high to low is: AB-8 macroporous resin purified>ethyl acetate extract>ethanol extract. The binding rate of AB-8 macroporous resin to sodium taurocholate, sodium glycocholate and sodium cholate was 88.2%, 73.2% and 75.8 %, respectively, and the binding ability was significantly higher than other flavonoid samples ($p<0.05$).

2.4.3 Correlation between the purity of total flavonoids and the ability to bind cholate

The correlation coefficients between the total flavonoid purity and the combined cholate capacity of the three samples are shown in Table 3.

Table 3. The correlation between the purity of total flavonoids and the ability to bind cholate.

| | Sodium glycocholate sodium binding rate | Sodium cholate binding rate | Sodium taurocholate binding rate |
|---|---|---|---|
| Total flavonoid purity | 0.963 | 0.983 | 0.988 |

From Table 3, the purity of total flavonoids was well correlated with the binding rate of sodium glycocholate, the binding rate of sodium cholate, and the binding rate of sodium taurocholate. The correlation coefficient was between 0.963 and 0.993. Cholesterol in the human body is metabolized and decomposed by synthesizing bile acids, excreting cholates, promoting the conversion of cholesterol in the liver into cholate, promoting cholesterol degradation and metabolism, thereby reducing blood fat [7]. Therefore, it is indicated that total flavonoids may be a factor in binding to cholate. The higher the purity of total flavonoids, the stronger the ability to bind cholates, indicating that the lowering of blood lipids is better [19].

## Conclusion

In this paper, the ethanol extract of Camellia sinensis was extracted by different polar solvents, and then purified by AB-8 macroporous resin to improve the purity of total flavonoids. The purity of total flavonoids in ethanol extract was 27.8%, the purity of total flavonoids in ethyl acetate extract was 46.4%, and the purity was increased by 18.6%. The static test of 8 different resins was carried out, and the AB-8 macroporous resin was selected as the most hanging resin for purifying the total flavonoids of the sunflower flower. The dynamic test conditions of AB-8 macroporous resin were optimized, and the total flavonoid concentration of the sample solution was determined to be 5.5 mg/mL, and the sample volume was 110 mL under the flow rate of 1.5 mL/min; 75% ethanol was used as the eluent at the flow rate. The elution amount of 80 mL at 1.5 mL/min increased the purity of total flavonoids from 46.4% to 83.5%, and increased by 37.1%, and obtained good purification results. The binding rate of AB-8 macroporous resin to sodium taurocholate, sodium glycocholate and sodium cholate was 88.2%, 73.2% and 75.8 %, respectively. The binding ability was the strongest, and the others were ethyl acetate extract. , ethanol extract. The correlation coefficient between total flavonoid purity and bile acid binding capacity in the sample ranged from 0.963 to 0.988, indicating that the total flavonoids of the sunflower flower had good hypolipidemic effect in vitro.

## Acknowledgments

This work was a supported by the National Key Research Development Program [Grant No. 2016YFD0400600].